\documentclass[11pt,a4paper,oneside,opernright]{article}
\usepackage[utf8]{inputenc}
\usepackage[T2A,T1]{fontenc}
\usepackage{titlesec}
\usepackage{amsmath}
\usepackage{amsfonts}
\usepackage{mathtools}
\usepackage{amssymb}
\usepackage{graphicx}
\usepackage{amsthm}
\usepackage{setspace}
\usepackage{fullpage}
\usepackage{rotating}
\rmfamily
\makeindex
\makeatletter
\titleformat{\section}
{\normalfont\large\bfseries}{\thesection}{1em}{}

\begin{document}
\title{Citation gaming induced by bibliometric evaluation: a country-level comparative analysis  \footnote{Funding: This work was supported by Institute For New Economic Thinking Grant ID INO17-00015.} }
\author{Alberto Baccini\footnote{Department of Economics and Statistics, University of Siena, Italy; alberto.baccini@unisi.it}\and Giuseppe De Nicolao\footnote{Department of Electrical, Computer and Biomedical Engineering, University of Pavia, Italy}\and Eugenio Petrovich\footnote{Department of Economics and Statistics, University of Siena, Italy} }
\date{}
\maketitle
\abstract It is several years since national research evaluation systems around the globe started making use of quantitative indicators to measure the performance of researchers. Nevertheless, the effects on these systems on the behavior of the evaluated researchers are still largely unknown.  We attempt to shed light on this topic by investigating how Italian researchers reacted to the introduction in 2011 of national regulations in which key passages of professional careers are governed by bibliometric indicators. A new inwardness measure, able to gauge the degree of scientific self-referentiality of a country, is defined as the proportion of citations coming from the country itself compared to the total number of citations gathered by the country. Compared to the trends of the other G10 countries in the period 2000-2016, Italy's inwardness shows a net increase after the introduction of the new evaluation rules. Indeed, globally and also for a large majority of the research fields, Italy became the European country with the highest inwardness. Possible explanations are proposed and discussed, concluding that the observed trends are strongly suggestive of a generalized strategic use of citations, both in the form of author self-citations and of citation clubs. We argue that the Italian case offers crucial insights on the constitutive effects of evaluation systems. As such, it could become a paradigmatic case in the debate about the use of indicators in science-policy contexts.
\\


\newpage

\section{Introduction}

Starting from the late 1980s, several European and extra-European countries implemented national systems to monitor, assess, and evaluate the research performance of their scientific workforce (1, 2). One of the key features of such research evaluation systems is the focus on quantitative indicators (metrics) as crucial science policy tools (3). Accordingly, in the last years, several scientometric indicators, based on publications or citations (or on a combination of both, such as the \textit{h-index}), have increasingly appeared in the academic evaluation systems, alongside with the traditional peer-review-based procedures. 

The use of these indicators in the evaluation of research performance has generated a heated debate in the scientific community. The advocates argue that scientometric measures are not only more objective than the peer-review (4); they would also improve both the quantity and the quality of the scientific production (5, 6). This would occur because the indicators are integrated within a system of incentives that rewards the achievement of the scientometric targets set by the evaluation system (7). On the other hand, critics claim that the same mechanisms that are designed to improve the research performance create at the same time room for strategic behaviors  (8). For instance, when productivity is positively rewarded, the number of publications become a goal that can be pursued not only by positive behaviors (doing more research), but also by opportunistic strategies (e.g., slicing one scientific work into multiple publications) (9, 10). Analogously, when citations become a goal, the «citation game» starts (11). A mediating position is represented by scholars proposing a «responsible use» of metrics. According to this approach, research metrics can provide valuable insights on the research performance, granted that they are carefully designed in order to avoid unintended consequences. Thus, a distillation of best practices has been proposed for improving the use of metrics in research assessment (12).

Recently, the idea that the consequences of the use of indicators on the behavior of researchers can be easily sorted between the intended  and the unintended  ones, has been questioned as too simplistic (13, 14).  Instead, the notion of «constitutive effects» has been advanced to capture the way in which the indicators act on the researchers (15). Within this new framework, indicators are conceived as shaping the activity of research deeply and at different levels, from the citation habits to the research agenda, redefining at the same time key evaluative terms such as research quality (16). They become crucial actors in the «epistemic living spaces» of academic researchers (17) and researchers begin to «think with indicators» pervasively (18).
\\
The main constitutive effects of the indicators described in the literature can be grouped into three main types: i) Goal-displacement: scoring high on the indicators becomes a target in itself, that is to be achieved also by \textit{gaming} the system (19, 20); ii) Risk avoidance: highly innovative, not mainstream, and interdisciplinary research topics are avoided because they could do not score well on indicators that tend to reward more traditional research programmes (18, 21–25); iii) Task reduction: when academic activities such as teaching and public engagement are not rewarded, academics tend to avoid them to concentrate only on publishable academic research (26–28).
\\
Although these effects have been highly debated, until recently the evidence of their occurrence has been mainly anecdotal. It is only in the last years that the methodical empirical study of such effects has been undertaken (13, 21). 
In the present paper, we aim to advance the knowledge on this topic by focusing on the case of Italy. Among European and extra-European countries, Italy is the only one in which some key career passages of scientific researchers \footnote{Except for the scholars in the Social Sciences and Humanities (see next section).} are entirely regulated by rules based on bibliometric indicators. Thus, Italy is ideally suited to studying the response of researchers to the use of metrics in research evaluation.

In particular, we will investigate whether Italian scientists have pervasively adopted a strategic use of citations in order to boost their indicators. By “pervasively”, we mean that the effect of this behavior should be visible \textit{in the great majority of scientific fields, at the national level}. As we will highlight in the Conclusion, the Italian case provides important insights on the constitutive effects of evaluation systems in general.

The rest of the paper is organized as follows. In the next two sections, the specificity of the Italian case is explained and the literature dealing with self-citing strategic behaviors is reviewed. Next, a new “inwardness” indicator is introduced that is sensitive to collective strategic citation behaviors at a country level. In the Data section, the procedure for retrieving the data is described, while the main findings are presented in the Results section. In the Discussion, after examining alternative explanations, it is argued in favor of the emergence of a collective strategic behavior devised to meet the demands of the evaluation system. In the Conclusion, some general lessons from the Italian case are drawn.

\section{The Italian case}

In 2010, the Italian university system underwent a wide process of reformation, regulated by the Law 240/2010. The reform created the Agency for the Evaluation of the University and Research (ANVUR), a centralized agency whose main task is the monitoring and the evaluation of the Italian research system. The Agency started in 2011 a research assessment exercise called VQR, relative to the period 2004-2010. A second research assessment exercise was started in 2015, relative to the period 2011-2014. In both exercises, the evaluation of submitted articles was largely based on the automatic or semi-automatic use of algorithms fed by citation indicators (29) while other research outputs, such as books, were evaluated by peer reviews.

The reform modified also the recruitment and advancement system for university professors by introducing the National Scientific Habilitation (ASN). Both for hiring and promotion, having obtained the ASN has become mandatory for applying to academic positions. The bibliometric rules rely on three indicators. For the hard sciences, life sciences, and engineering, the indicators considered by ANVUR are the number of journal articles, the number of citations, and the h-index.  For the social science and humanities, the indicators are the number of research outputs, the number of monographs, and the number of papers published in “class A” journals. At each new round of habilitation, ANVUR calculates for each of these indicators the “bibliometric thresholds” that the candidates must overcome to achieve the ASN.\footnote{For the first edition of the ASN the national rules were defined in the Ministerial Decree 7 June 2012 n. 76. http://attiministeriali.miur.it/media/192901/dm\_07\_06\_12\_regolamento\_abilitazione.pdf. ANVUR defined the thresholds used for the first edition of the ASN: https://web.archive.org/web/20190207112821/http://www.anvur.it/attivita/asn/asn-2012-2013/indicatori-e-relative-mediane/.}
Candidates whose indicators do not overcome two thresholds out of three cannot be habilitated (exceptions were possible in specific circumstances only in the first edition, ASN 2012). When first introduced, the thresholds were stated to be the median values of the indicators of the permanent academic staff holding that position (associate or full professor). To make and example, in order to obtain a full professor habilitation, the candidate was required to score better than half of the current full professors in two indicators out of three.  Applicants overcoming the fixed thresholds are then evaluated by a committee composed by five referees who are in charge of the final decision about attributing habilitation.

Note that the focus on indicators is not confined to the national procedures but “trickles down” to the university committees in charge of recruiting and promotion that are required to take into account production and citation metrics when they evaluate and rank the habilitated applicants. Finally, also the members of both the national habilitation and the local recruitment committees are required to overcome bibliometric thresholds. 

In sum, in Italy, starting from 2011, bibliometric indicators have gained a central role not only in the national research assessment but in the entire body of the recruitment procedures. 
A remarkable peculiarity of the Italian system is that the indicators based on citations, used both in the habilitation procedure and in the research evaluation exercise, are calculated \textit{by including self-citations}. Thus, researchers can increase their indicators just by self-citing their own work.

Anecdotal evidence of the adoption of strategic behaviors in the form of author self-citations has been presented by Baccini (30). Two recent studies have documented more thoroughly the rise of opportunistic behaviors in response to the ASN rules. Seeber et al. has analyzed how the use of self-citations in four Italian research areas changed after the introduction of the habilitation procedure. They have found that scientists in need of meeting the thresholds (i.e., those looking for habilitation as a prerequisite for tenure-track or promotion to full professor) did increase significantly their self-citations after 2010 (31). Scarpa et al. focused on the Italian engineering area and found an anomalous peak in the self-citations rate (i.e., the number of self-citations to the total number of citations) in correspondence of the second round of the habilitation procedure, in 2013. (32). 

\section{Strategic behaviors and country self-citations}

Even if the afore-mentioned studies have highlighted some recent behavior changes of Italian scientists, they did not address a subtler form of strategic behavior, the one based on the so-called «citation clubs» or «citation cartels». A citation club is an informal structure in which citations are strategically exchanged among its members to boost the respective citation scores (33–35). Note that this kind of strategy cannot be spotted when we use the standard definition of self-citation, according to which a self-citation occurs whenever the set of co-authors of the citing papers and that of the cited one are not disjoint (36, 37), because the members of the citation club might not be also co-authors. 
In order to allow for the effects of citation clubs, we examine a particular – and not much studied – form of self-citations, namely the \textit{country self-citations} (38). A country self-citation occurs whenever the set of the countries of the authors of the citing publication and the set of the countries of the authors of the cited publication are not disjoint, that is, if these two sets share at least one country (39, 40). Notably, any citation exchanged within a citation club formed by researchers working in the same country is counted as country self-citations, even when it is not an author self-citation. 

Thus, considering that most of the standard author self-citations are country self-citations too,\footnote{The only exception being authors that changed their country between the citing and the cited publication.} by analyzing the country self-citations, we can capture both the “classic” strategy based on author self-citations, and the “elaborated” one based on citation clubs.  As far as we are interested in countries and not in the individual authors, we will say, by short, that a paper is “authored by a country” when at least one of its authors is from that country.

Just as not all author self-citations originate from gaming purposes, in the same way not all country self-citations are the result of opportunistic behaviors. Indeed, the literature on author self-citations agrees on the fact that a certain amount of them is a normal byproduct of the scientific communication. There are many perfectly legitimate reasons for citing one’s own works, such as building on previously obtained results, avoiding repetition, and so on (41–43). By the same token, it is normal that a country has an internal exchange of citations amongst its researchers insofar the knowledge produced by the country is used (i.e., cited) by the same country's scientific staff.

Moreover, international collaboration positively affects the number of country self-citations. In fact, the more a country collaborates with other countries, the higher will be the number of country self-citations. Take for instance a paper authored in collaboration by Italy and France. Any future citation to that paper coming from an Italian-authored or a French-authored publication will count as a country self-citation for both Italy and France, since the citing and the cited publication will share at least one country of affiliation. 

In sum, the country self-citations are not \textit{per se} a sign of strategic behavior since they depend both on the internal exchange of knowledge within a country and the amount of international collaboration. Nonetheless, if the researchers of a single country initiate strategic behaviors in order to boost their citations, this is likely to produce an \textit{anomalous increase} of country self-citations compared to the other countries.

\section{The inwardness indicator}
In order to obtain a normalized measure of country self-citations, we introduce a simple indicator of “inwardness”. For a given year and a country $c$, the inwardness is defined as the percentage ratio between the total number of country self-citations ($S_c$) and the total number of citations ($C_c$) of that country :
\begin{equation}
I_c=\frac{S_c}{C_c}\times100
\end{equation}
The minimum value of the inwardness indicator is $I_c=0$ when a country has no self-citations; and the maximum is $I_c=100$ when a country has self-citations only, that is $S_c=C_c$.It is easy to show that the inwardness indicator is a variant of the Relative Citation Impact ($RCI$) of a country. The $RCI$ is defined by May (44) as the ratio between the average citation per paper of a country and the average citation per paper of the world (see also (45)). The $RCI$ of the country $c$ in a given year is defined as $RCI_c=\dfrac{C_c}{P_c}\times \dfrac{P_w}{C_w}$ where $C_c$ and $C_w$ are the total number of citations of the country and of the world, and $P_c$ and $P_w$ the publications of the country and of the world. The total number of citations is the sum of the country self-citations ($S_c$) and the external citation ($X_c$); when the world is considered $C_w=S_w$, since obviously $X_w=0$. If a Relative Self-citation Impact is defined as $RSI_c=\dfrac{S_c}{P_c}\times \dfrac{P_w}{S_w}$, the inwardness indicator can be expressed as\\
\begin{equation}
I_c=\dfrac{RSI_c}{RCI_c}=\left(\dfrac{\dfrac{S_c}{P_c}}{\dfrac{S_w}{P_w}}\right)\times \left(\dfrac{\dfrac{C_w}{P_w}}{\dfrac{C_c}{P_c}}
\right)=\dfrac{S_c}{C_c}
\end{equation}
Note that the inwardness indicator is normalized for the size of the country in terms of publications. 
From a conceptual point of view, the inwardness of a country  is an indicator of how much the knowledge produced in the form of scientific publications in a given year in a country flows, through citations, into the knowledge produced in that country in the following years (46–48). Indeed, $1-I_c$  indicates how much of the knowledge produced in a year in a country flows, through citations, into the knowledge (publications) produced by other countries (49, 50). A higher level of inwardness suggests that the knowledge produced by a country attracts mainly the interest of the national community. By contrast, a lower level suggests that the research of the country does not remain confined within its own borders but flows also toward the rest of the world.

As said above, the strategic use of citations, both as author self-citations and as citation clubs, affects the country self-citations and, hence, also the inwardness indicator. The start of a strategic use of citations at the country level should therefore be associated with an \textit{anomalous} rise of the inwardness indicator.

Recall, however, that inwardness is positively affected also by increases of international collaboration. It is therefore necessary to control the trend of the international collaboration before concluding that an inwardness rise is due to strategic behaviors and not to an increase of international collaboration.
\section{Data}
We retrieved the data for calculating the Inwardness indicator from SCIval, an Elsevier’s owned platform powered by Scopus data. (https://www.scival.com/home)\footnote{The data were exported from SCIval on October 16, 2018. They correspond to the last update on Scopus of September 21, 2018.} In particular, we exported from SCIval two metrics: (1) Citation Count including self-citations, and (2) Citation Count excluding self-citations. For both metrics, we included articles, reviews, and conference papers, leaving aside other types of publications. The first Citation Count metrics represents the countries’ total number of citations, whereas the countries’ number of self-citations was obtained as the difference between (1) and (2).
\footnote{Note that the SCIval’s definition is binary and non-fractional: a citation can either be a self-citation or not (51). The weight of a country self-citation remains always 1, irrespective of the number of countries producing the citing or the cited publications: if an Italian publication is cited by another Italian publication, this self-citation will have the same weight as if the same publication was cited by an international Italo-French-Chinese publication.} 

We retrieved the data for the G10 countries (Belgium-BE, Canada-CA, France-FR, Germany-DE, Italy-IT, Japan-JP, the Netherlands-NL, Sweden-SE, Switzerland-CH, United Kingdom-GB, United States-US). In order to study the spread of the strategic behavior in different research areas, data were exported for all the Scopus fields aggregated, i.e., without any filter for subject area, and for each of the 27 Scopus Main Categories (total number of datasets = 28), for the years 2000-2016 included. 
In order to account for the effect of international collaboration on the inwardness indicators, we retrieved from SCIval also the Percentage of International Collaboration metric for the target countries. The percentage of international collaboration for a country in a given year is defined as the share of publications of the country coauthored by at least one different country. The graphs were implemented in R by using the package “ggplot2” (52). 

\section{Results} 
Figure 1 shows the trend of the inwardness over time for the eleven target countries (all Scopus fields aggregated). All countries share a rather similar profile with apparent differences in the absolute value. The ranking is partially explained by the size of the scientific production of the countries. Countries with a large scientific output, such as the Unites States, naturally attract more citations from their own production, simply because they have more citing and citable articles than smaller countries such as Belgium. For all the countries under analysis, not only the inwardness increases slowly and regularly over time, but the yearly ranks of countries according to their inwardness are remarkably stable.

\begin {figure} 
\centering
\includegraphics[scale=0.9]{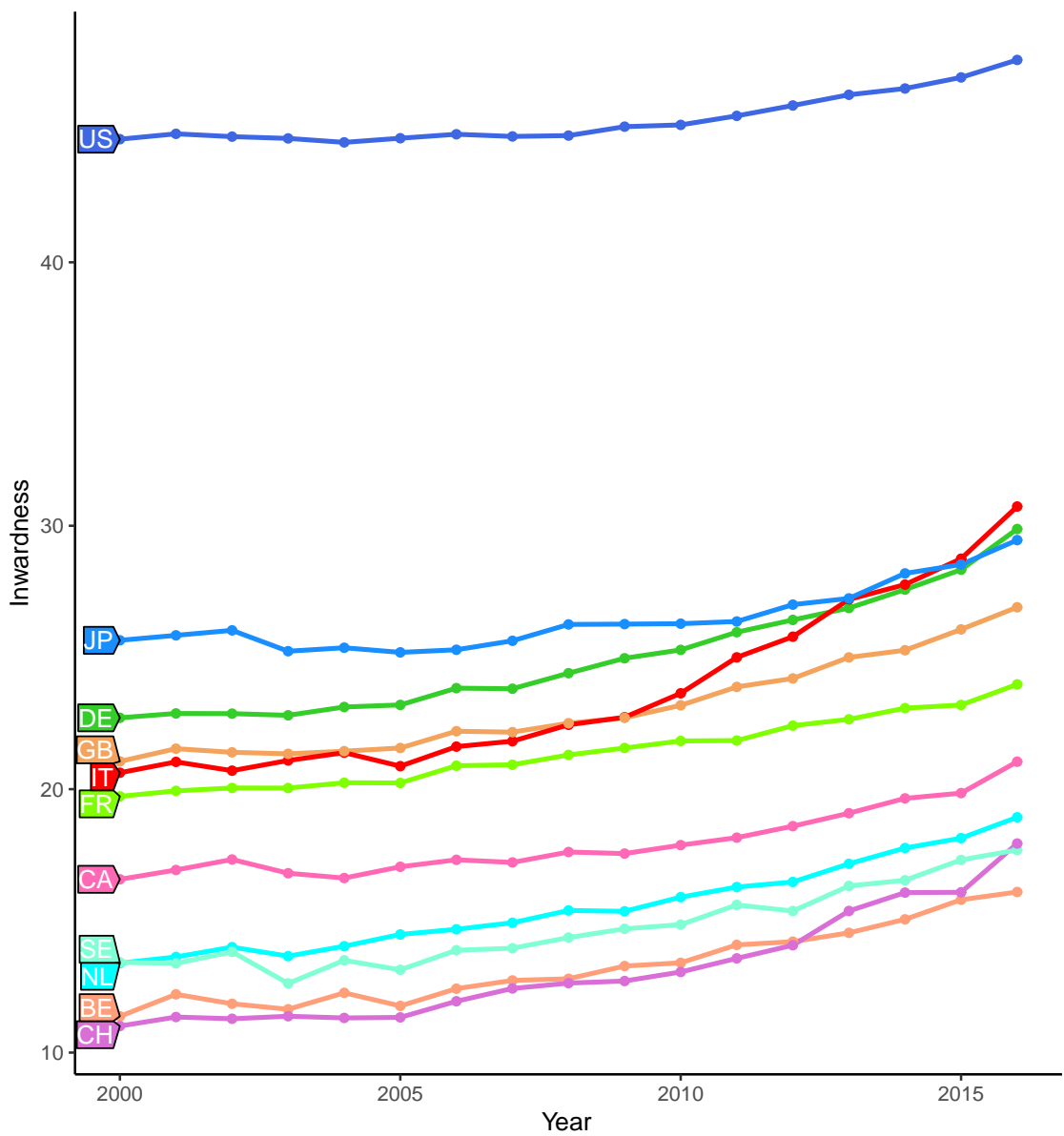} 
\caption { Inwardness for G10 countries (2000-2016). Source: elaboration on SCIval data.} 
\end {figure} 

In this landscape, Italy stands out as a notable exception. In 2000, at the beginning of the period, Italy has an inwardness of 20.62\% and ranks sixth, just behind UK. In 2016, at the end of the period, Italy ranks second, with an inwardness of 30.73\%. Note that, until 2009, Italy’s inwardness grows parallel to those of comparable countries (UK, Germany, France). However, around 2010, the Italian trend shows a sudden acceleration. In the following six years, Italy overcomes UK, Germany, and Japan, becoming the first European country and the second one in the G10 group.

Table 1 shows the variations (deltas) of the inwardness for each country, for the whole period and by considering two sub-periods, 2008-2000 and 2016-2008. Note that in the first period, Italy’s increase is in line with other countries, while in the second period (2008-2016), Italy’s exhibits the largest inwardness delta: 8.29 p.p., more than 4 p.p. above the G10 average and almost 3 p.p. above Germany. As a result, Italy is by far the country with the highest inwardness delta also in the whole period 2000-2016 (10.11 p.p. vs 5.22 of the G10 average).

\begin{table}
\centering
\caption{Inwardness delta. Delta is calculated as simple difference (p.p.) between the inwardness in the last and the first year of the period.}
\begin{tabular}{cccc}
\hline 
Country& $\Delta_1$ (2000-2008)& $\Delta_2$ (2008-2016)&$\Delta_{tot}$ (2000-2016)\\ 
\hline 
Belgium&1.42&3.29&4.72 \\ 
\
Canada&1.04&3.43&4.46 \\ 
\
France&1.57&2.68&4.25 \\ 
\
Germany&1.69&5.47&7.17 \\ 
\
Italy&1.82&8.29&10.11 \\ 
\
Japan&0.6&3.2&3.81 \\ 
\
Netherlands&2&3.54&5.54 \\ 
\
Sweden&0.94&3.32&4.27 \\ 
\
Switzerland&0.94&3.32&4.27 \\ 
\
United Kingdom&1.45&4.4&5.85 \\ 
\
United States&0.14&2.87&3.01 \\ 
\
\textit{Mean G10 countries}&1.24&3.98&5.22 \\ 
\hline 
\end{tabular} \\
\end{table}
\
However, as already said, inwardness is affected by the amount of International Collaboration of a country. In order to allow for this effect, in Figure 2, inwardness is plotted against the average international collaboration score of each country. More precisely, inwardness at year Y is plotted against the three-years moving average value of international collaboration calculated starting from year Y. In fact inwardness at year Y depends also on citations coming from publications appeared in the following years (53).

The data shows indeed a positive relation between the two variables: for all the countries, inwardness grows with the average international collaboration. The plot shows a peculiar trajectory for Italy. Although for most years Italy ranks last in Europe for international collaboration (x-axis), nevertheless, at the end of the period, it is the first European country for inwardness (y-axis). Before 2010, Italy is close to and moves together with a group of three European countries, namely Germany, UK, and France. Starting from 2010, Italy departs from the group along a steep trajectory, to eventually become the European country \textit{with the lowest international collaboration and the highest inwardness}.

\begin {figure} 
\centering
\includegraphics[scale=0.9]{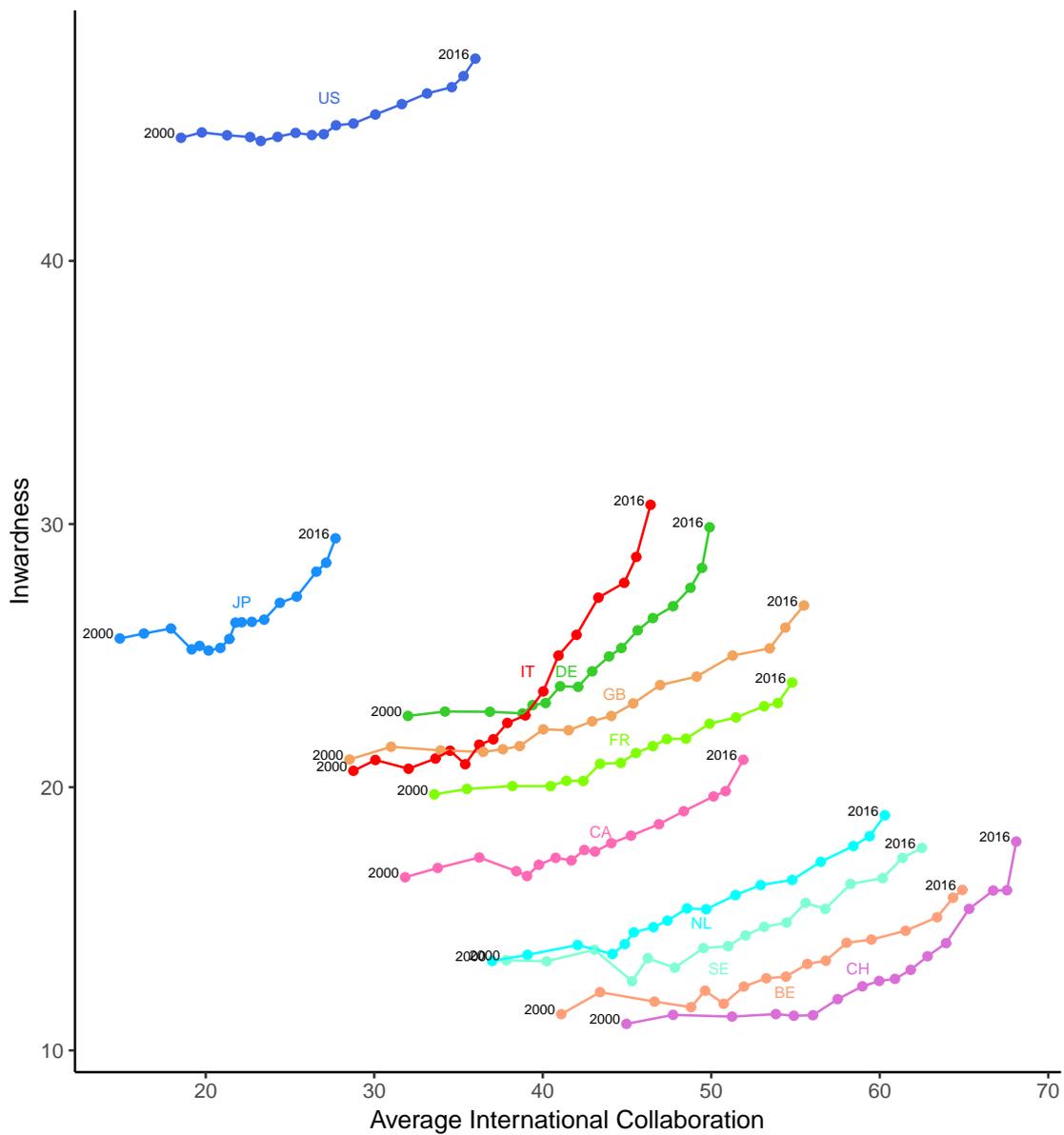} 
\caption { Inwardness versus average international collaboration for the G10 countries. The average international collaboration is the 3-year moving average calculated starting from the considered year. The international collaboration is defined as the share of publications of a country coauthored by at least a coauthor of a different country. Source: elaboration from SCIval data.} 
\end {figure} 
Until now, we focused on the aggregated output of the target countries, without considering the different research areas (Scopus Main Categories). In order to investigate whether and how inwardness changes across research areas, we calculated the inwardness time series for each of the 27 Scopus Main Categories. The time series, as well as the scatterplots of the inwardness against the international collaboration, are fully provided in the Supplementary Materials. For reasons of space, these data are summarized in Figure 3, where the variation of the inwardness indicator in the periods 2000-2008 (A) and 2008-2016 (B) is displayed for each of the 27 Scopus Categories.
Italy shows a remarkable difference between the two periods. In the first one (Figure 3A), before the university reform, Italy is in line with the other G10 countries in most of the research fields. In the second period, after the reform, Italy stands out with the highest inwardness increase in 23 out of 27 fields. The only exceptions are earth and planetary sciences (EPS), multidisciplinary (MUL), nursing (NUR), and physics and astronomy (PA). 

\begin {figure} 
\centering
\includegraphics[scale=0.5]{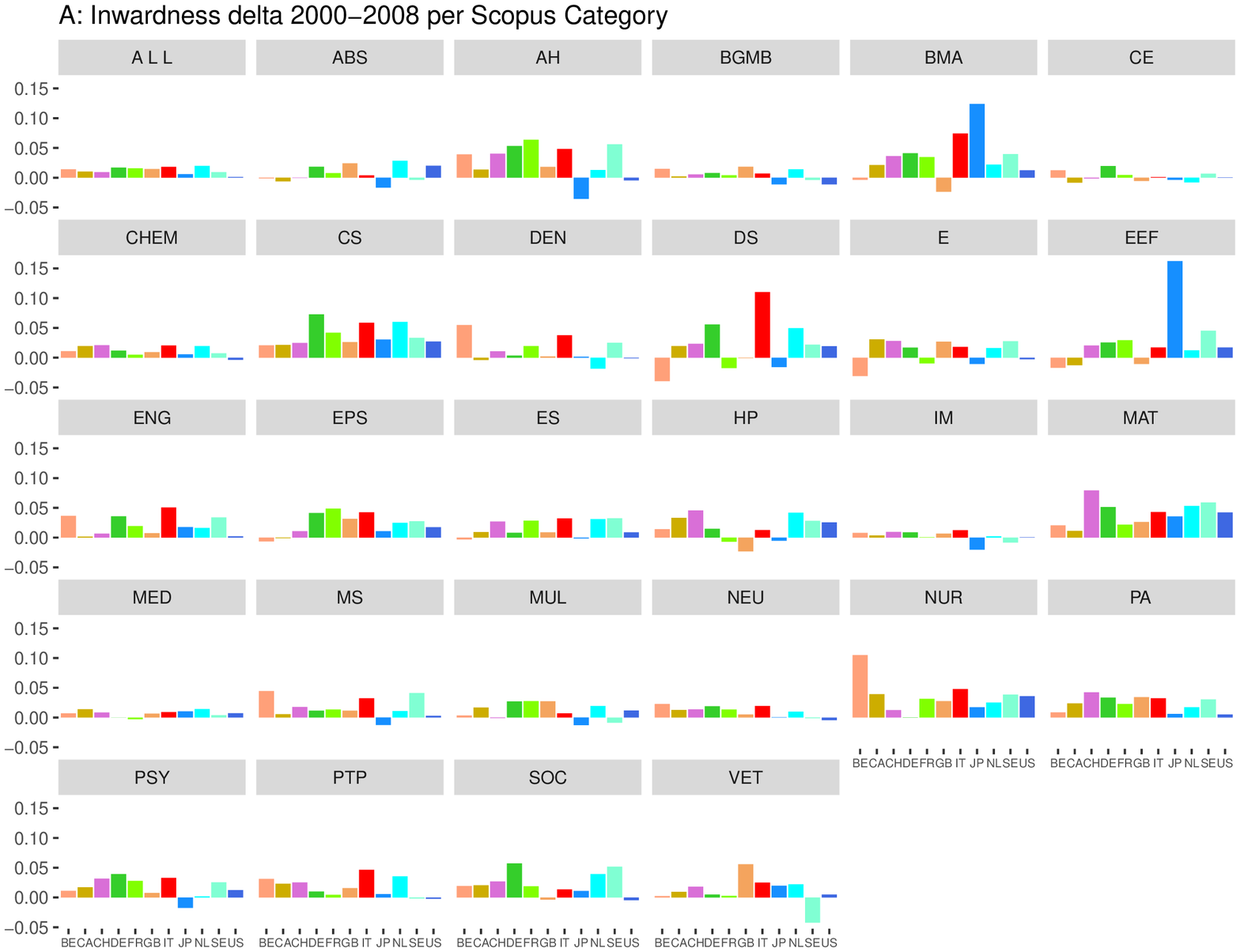} 
\includegraphics[scale=0.5]{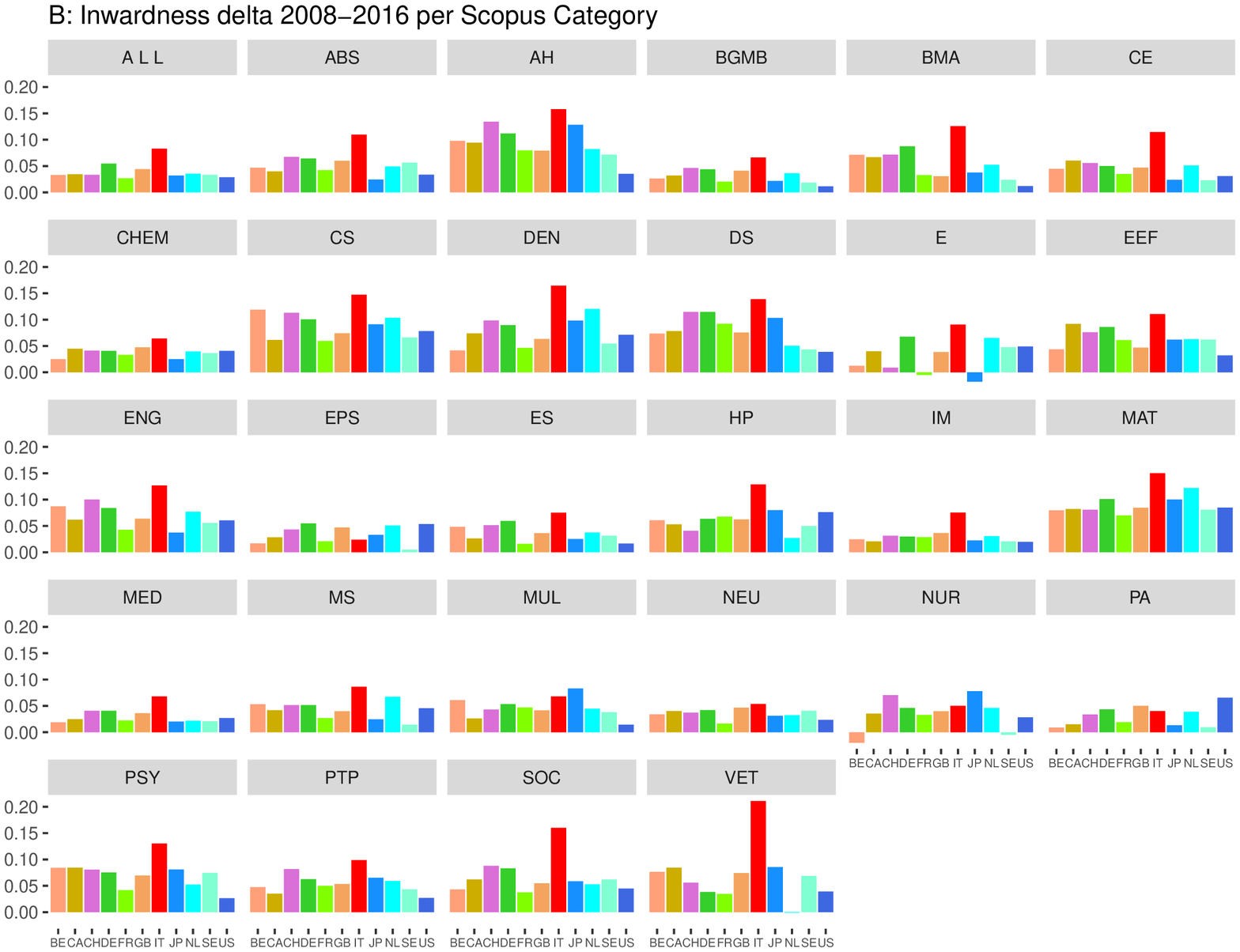} 
\caption {Inwardness delta in Scopus Main Categories in the periods 2000-2008 (A) and 2008-2016 (B). ABS = Agricultural and Biological Sciences, AH = Arts and Humanities, BGMB = Biochemistry, Genetics and Molecular Biology, BMA = Business, Management and Accounting, CE = Chemical Engineering, CHEM = Chemistry, CS = Computer Science, DEN = Dentistry, DS = Decision Sciences, E = Energy, EEF = Economics, Econometrics and Finance, ENG = Engineering, EPS = Earth and Planetary Sciences, ES = Environmental Science, HP = Health Professions, IM = Immunology and Microbiology, MAT = Mathematics, MED= Medicine, MS = Materials Science, MUL = Multidisciplinary, NEU = Neuroscience, NUR = Nursing, PA = Physics and Astronomy, PSY = Psychology, PTP = Pharmacology, Toxicology and Pharmaceutics, SOC = Social Sciences, VET =  Veterinary. Source: elaboration from SCIval data.}
\end {figure} 

As we show in the Supplementary Materials, the inwardness increase is \textit{not} matched by a parallel increase of the international collaboration at the field level. In particular, at the end of the period, Italy is the European country with the lowest level of international collaboration and the highest value of inwardness in the following Scopus Categories (11 on 27): agricultural and biological sciences (ABS), biochemistry, genetics and molecular biology (BGMB), chemical engineering (CE), economics, econometrics and finance (EEF), earth and planetary sciences (EPS), environmental science (ES), immunology and microbiology (IM), pharmacology, toxicology and pharmaceutics (PTP),  veterinary (VET). In other 9 Categories, Italy is first for inwardness but not the lowest for international collaboration: business, management and accounting (BMA), computer science (CS), dentistry (DEN), decision sciences (DS), engineering (ENG), health professions (HP), mathematics (MAT), materials science (MS), psychology (PSY). Note that the Italian production in the arts and humanities (AH) and social sciences (SOC) is only partially covered by Scopus as a large part is published in books and in the national language. Therefore, the results about these scholarly areas should be taken with great caution (54).
\section{Discussion}
As seen from Figure 1 and Table 1 Italy shows a different trend compared to the other G10 countries. The most notable aspect is that, after 2009, Italy's inwardness \textit{grows faster}. The acceleration is about synchronous with the launch of the national assessment exercise in 2011 and the opening in 2012 of the ASN, the new scientific habilitation system, whose bibliometric criteria, largely relying on citations, had been announced in 2011. 
A likely explanation of the anomalous trend is that the Italian scientific community reacted to the bibliometric thresholds set by ANVUR by citing more frequently the Italian scientific production. More specifically, we argue that the change in the citation behavior is due to the widely spread adoption, by Italian researchers, of strategies for boosting bibliometric indicators set by ANVUR.  As said in the Introduction, such strategies include both the artificial increase of author self-citations and the creation of nationally-based citation clubs. The Italian anomalous trend is possibly the joint result of these two strategic answers to the incentives of the evaluation system.
The slight discrepancy between the starting year of the inwardness acceleration and the launch of bibliometric evaluation system, with the former occurring slightly earlier than the latter, is easily explained by the “backward effect” typical of citation measures.  Any change in the citation habits taking place in a given year produces a backward effect on the citation scores of the previous years because researchers cite previously published papers, so that the change reverberates also on the citation scores of the past production. The Italian ASN used time horizons of 10 and 15 years for counting citations and for calculating h-indexes of applicants and referees. Citations received by the most recent articles have a more lasting effect in the calculations of forthcoming indicators. It is therefore more convenient to self-cite one's own recent production rather than the remote one. Hence, a strategic reaction to rules introduced in year 2011 is expected to produce an inwardness acceleration that starts a few years before, just as observed for Italy.

Two alternative explanations of the data could be advanced. The Italian acceleration may be due to a sudden rise, after 2009, of the amount of international collaborations. In fact, we have already observed that, other things left unchanged, an increase of international collaboration positively affects the inwardness indicator. However, Figure 2 rules out this alternative explanation. No peculiar increase in the Italian international collaboration can be spotted. 
A second alternative explanation argues that the inwardness acceleration may due to the narrowing of the scientific focus of Italian researchers, i.e. to a dynamic of scientific specialization  which led to a growth of author self-citations (31). The idea is that focusing on narrower topics results in a contraction of the scientific community of reference. Thus, the number of citable papers would diminish and the chances for author self-citation would correspondingly increase, generating also the growth of the country self-citations. Actually, no evidence can be showed that directly falsifies the specialization explanation. Nonetheless, this explanation appears implausible because it would imply that Italian researchers in all fields have suddenly redirected their focus to topics mainly investigated in the national community. This changing behavior would be not only peculiar of Italy, but also so strong to lead Italy to diverge from the other G10 countries in terms of inwardness. Notably, Figure 3 shows that the post-2008 acceleration is visible in most of the research areas in Italy. Not only the change in the behavior has been generalized, regarding most of the fields of research, but in some fields, such as engineering (ENG), mathematics (MAT) or veterinary (VET), the increase reached outstanding proportions. In any case, it would still be necessary to explain why specialization occurred only in Italy and at the same time as the adoption of new rules for evaluation.  

\section{Conclusion}
In this paper, we contributed to the empirical study of the constitutive effects that indicator-based research evaluation systems have on the behavior of the evaluated researchers. By focusing on the Italian case, we investigated how the Italian scientific community responded, \textit{at the national level}, to the introduction of a research evaluation system, in which bibliometric indicators play a crucial role. 
Our results show that the behavior of Italian researchers has indeed changed after the introduction of the evaluation system following the 2010 university reform. Such a change is visible \textit{at a national scale in most of the scientific fields}. We explained this as the result of the pervasively adoption of strategic citation behaviors within the Italian scientific community. In particular, the inwardness indicator was able to track the effects of two types of citation strategies: the opportunistic use of author self-citation and the creation of citation clubs exchanging citations between their members. Even if further research is needed to assess the respective weight of these two strategies, it is their joint presence that best explains the peculiar trend of the Italian inwardness, exhibiting a neat acceleration after 2010. 

In sum, the comparative analysis of the inwardness indicator showed that Italian research grew in insularity in the years after the adoption of the new rules of evaluation. Indeed, results show that, both globally and for many research fields, while the level of international collaboration remained stable and comparatively low, the research produced in the country tended to be increasingly cited by papers authored by at least an Italian scholar. Put in other words: the share of citations to Italian articles received by articles authored by non-Italian authors sharply decreased after 2010.

We believe that three main lessons can be derived from the Italian case. Firstly, our results confirm that scientists are \textit{quickly responsive} to the system of incentives in which they act (31). Thus, any policy aiming at introducing or modifying such a system should be designed and implemented very carefully. In particular, considerable attention should be placed on the constitutive effects of bibliometric indicators. They are not neutral measures of performance but actively interact and quickly shape the behavior of the evaluated researchers.

Secondly, our results show that the «responsible use» of metrics would not be enough to prevent the emergence of strategic behaviors. For instance, the Leiden Manifesto recommends the use of a «suite of indicators» instead of a single one as a way to prevent gaming and goal displacement (see the principle number 9 in (12)). The Italian case shows that, even if the researchers are evaluated against multiple indicators, as recommended, strategic behaviors manifest themselves anyway. 

Lastly, our results prompt some reflections on the viability of the mixed evaluation systems, in which the indicators are intended for complementing or integrating the expert judgment expressed by the peer review. In fact, the Italian system was designed in principle according to such a mixed approach, both for the research assessment exercises where research products were evaluated by bibliometric indicators or by peer reviewers, and for the ASN where to overcome bibliometric thresholds is but a necessary condition for being admitted to the final evaluation by habilitation committees. Nonetheless, our results show that the mere presence of bibliometric indicators in the evaluative procedures is enough to structurally affect the behavior of the scientists, fostering opportunistic strategies. Therefore, there is the concrete risk that in mixed evaluation systems, the indicator-based component overcomes the peer review-based one. Hence, they \textit{de facto} collapse to indicator-centric approaches. We believe that further research is needed to better understand and fully appreciate the possibility of such a collapse. In the meantime, we suggest that policy makers should exercise the most extreme caution in the use of indicators in science policy contexts.

\section*{References}
1. 	Hicks D (2012) Performance-based university research funding systems. \textit{Research Policy} 41(2):251–261.
\\
\\
2. 	Whitley R, Gläser J eds. (2007) \textit{The changing governance of the sciences: the advent of research evaluation systems} (Springer, Dordrecht, the Netherlands).
\\
\\
3. 	Haustein S, Larivière V (2015) The Use of Bibliometrics for Assessing Research: Possibilities, Limitations and Adverse Effects. \textit{Incentives and Performance}, eds Welpe IM, Wollersheim J, Ringelhan S, Osterloh M (Springer International Publishing, Cham), pp 121–139.
\\
\\
4. 	Moed HF (2005) \textit{Citation analysis in research evaluation} (Springer, Dordrecht).\\
\\
5. 	Geuna A, Martin BR (2003) University Research Evaluation and Funding: An International Comparison. \textit{Minerva} 41(4):277–304.\\
\\
6. 	Ingwersen P, Larsen B (2014) Influence of a performance indicator on Danish research production and citation impact 2000–12. \textit{Scientometrics} 101(2):1325–1344.\\
\\
7. 	Hicks D (2010) \textit{Overview of models of performance-based research funding systems. Performance-Based Funding for Public Research in Tertiary Education Institutions} (OECD), pp 23–52.\\
\\
8. 	Edwards MA, Roy S (2017) Academic Research in the 21st Century: Maintaining Scientific Integrity in a Climate of Perverse Incentives and Hypercompetition. \textit{Environmental Engineering Science} 34(1):51–61.\\
\\
9. 	Butler L (2003) Modifying publication practices in response to funding formulas. \textit{Research Evaluation} 12(1):39–46.\\
\\
10. 	Butler L (2005) What happens when funding is linked to publication counts? \textit{Handbook of Quantitative Science and Technology Research}, eds Moed HF, Glänzel W, Schmoch U (Springer, Dordrecht), pp 389–405.\\
\\
11. 	Biagioli M (2016) Watch out for cheats in citation game. \textit{Nature} 535(7611):201–201.\\
\\
12. 	Hicks D, Wouters P, Waltman L, de Rijcke S, Rafols I (2015) Bibliometrics: The Leiden Manifesto for research metrics. \textit{Nature} 520(7548):429–431.\\
\\
13. 	Rijcke S de, Wouters PF, Rushforth AD, Franssen TP, Hammarfelt B (2016) Evaluation practices and effects of indicator use—a literature review. \textit{Research Evaluation} 25(2):161–169.\\
\\
14. 	Wouters P (2018) The failure of a paradigm. \textit{Journal of Informetrics} 12(2):534–540.\\
\\
15. 	Dahler-Larsen P (2014) Constitutive Effects of Performance Indicators: Getting beyond unintended consequences. \textit{Public Management Review} 16(7):969–986.\\
\\
16. 	Biagioli M (2018) Quality to Impact, Text to Metadata: Publication and Evaluation in the Age of Metrics. \textit{KNOW: A Journal on the Formation of Knowledge} 2(2):249–275.\\
\\
17. 	Felt U, Červinková A (2009) \textit{Knowing and living in academic research: convergences and heterogeneity in research cultures in the European context} (Institute of Sociology of the Academy of Sciences of the Czech Republic, Prague).\\
\\
18. 	Müller R, de Rijcke S (2017) Thinking with indicators. Exploring the epistemic impacts of academic performance indicators in the life sciences. \textit{Research Evaluation} 26(3):157–168.\\
\\
19. 	Hammarfelt B, de Rijcke S (2015) Accountability in context: effects of research evaluation systems on publication practices, disciplinary norms, and individual working routines in the faculty of Arts at Uppsala University. \textit{Research Evaluation} 24(1):63–77.
\\
\\
20. 	Sousa SB, Brennan JL (2014) The UK Research Excellence Framework and the Transformation of Research Production. \textit{Reforming Higher Education}, eds Musselin C, Teixeira PN (Springer Netherlands, Dordrecht), pp 65–80.\\
\\
21. 	Fochler M, Felt U, Müller R (2016) Unsustainable Growth, Hyper-Competition, and Worth in Life Science Research: Narrowing Evaluative Repertoires in Doctoral and Postdoctoral Scientists’ Work and Lives. \textit{Minerva} 54(2):175–200.\\
\\
22. 	Gillies D (2008) \textit{How should research be organised?} (College Publications, London).\\
\\
23. 	Laudel G, Gläser J (2014) Beyond breakthrough research: Epistemic properties of research and their consequences for research funding. \textit{Research Policy} 43(7):1204–1216.\\
\\
24. 	Lee FS, Pham X, Gu G (2013) The UK Research Assessment Exercise and the narrowing of UK economics. \textit{Cambridge Journal of Economics} 37(4):693–717.\\
\\
25. 	Viola M (2018) Evaluation of Research(ers) and its Threat to Epistemic Pluralisms. \textit{European journal of analytic philosophy} 13(2):55–78.\\
\\
26. 	Broz L, Stöckelová T (2018) The culture of orphaned texts: Academic books in a performance-based evaluation system. \textit{Aslib Journal of Information Management} 70(6):623–642.\\
\\
27. 	van Dalen HP, Henkens K (2012) Intended and unintended consequences of a publish-or-perish culture: A worldwide survey. \textit{Journal of the American Society for Information Science and Technology} 63(7):1282–1293.\\
\\
28. 	Wilson M, Holligan C (2013) Performativity, work-related emotions and collective research identities in UK university education departments: an exploratory study. \textit{Cambridge Journal of Education} 43(2):223–241.\\
\\
29. 	Baccini A, De Nicolao G (2016) Do they agree? Bibliometric evaluation versus informed peer review in the Italian research assessment exercise. \textit{Scientometrics} 108(3):1651–1671.\\
\\
30. 	Baccini A (2018) \textit{Performance-based incentives, research evaluation systems and the trickle-down of bad science} (INET - Institute for New Economic Thinking, New York) 
https://www.ineteconomics.org/uploads/papers/Baccini-Value-for-money-Berlin-final.pdf.\\
\\
31. 	Seeber M, Cattaneo M, Meoli M, Malighetti P (2019) Self-citations as strategic response to the use of metrics for career decisions. \textit{Research Policy} 48(2):478–491.\\
\\
32. 	Scarpa F, Bianco V, Tagliafico LA (2018) The impact of the national assessment exercises on self-citation rate and publication venue: an empirical investigation on the engineering academic sector in Italy. \textit{Scientometrics} 117(2):997–1022.\\
\\
33. 	Šipka P (2012) Legitimacy of citations in predatory publishing: The case of proliferation of papers by Serbian authors in two Bosnian WoS-indexed journals. \textit{CEES Occasional Paper Series} (2012-12–2). Available at: http://www.ceon.rs/ops/12122.\\
\\
34. 	Van Noorden R (2013) Brazilian citation scheme outed. \textit{Nature} 500(7464):510–511.\\
\\
35. 	Fister I, Fister I, Perc M (2016) Toward the Discovery of Citation Cartels in Citation Networks. \textit{Frontiers in Physics} 4. doi:10.3389/fphy.2016.00049.\\
\\
36. 	Glänzel W, Bart T, Balázs S (2004) A bibliometric approach to the role of author self-citations in scientific communication. \textit{Scientometrics} 59(1):63–77.\\
\\
37. 	Snyder H, Bonzi S (1998) Patterns of self-citation across disciplines (1980-1989). \textit{Journal of Information Science} 24(6):431–435.\\
\\
38. 	Eto H (2003) Interdisciplinary information input and output of a nano-technology project. \textit{Scientometrics} 58(1):5–33.\\
\\
39. 	Elsevier (2018) \textit{Research Metrics Guidebook}. Available at: https://tinyurl.com/y2lq8f63
\\
\\
40. 	Tagliacozzo R (1977) Self-Citations in Scientific Literature. \textit{Journal of Documentation} 33(4):251–265.\\
\\
41. 	Pichappan P, Sarasvady S (2002) The other side of the coin: The intricacies of author self-citations. \textit{Scientometrics} 54(2):285–290.\\
\\
42. 	Garfield E (1979) Is citation analysis a legitimate evaluation tool? \textit{Scientometrics} 1(4):359–375.\\
\\
43. 	Hyland K (2003) Self-citation and self-reference: Credibility and promotion in academic publication. \textit{Journal of the American Society for Information Science and Technology} 54(3):251–259.\\
\\
44. 	May RM (1997) The Scientific Wealth of Nations. \textit{Science} 275(5301):793–796.\\
\\
45. 	Katz JS (2000) Scale-independent indicators and research evaluation. \textit{Science and Public Policy} 27(1):23–36.\\
\\
46. 	Merton RK (1974) \textit{The sociology of science: theoretical and empirical investigations }(Univ. of Chicago Pr, Chicago). \\
\\
47. 	Kaplan N (1965) The norms of citation behavior. Prolegomena to the footnote. \textit{American Documentation} 16(3):179–187.\\
\\
48. 	Zuckerman H (1987) Citation analysis and the complex problem of intellectual influence. \textit{Scientometrics} 12(5–6):329–338.\\
\\
49. 	Leydesdorff L (2007) Visualization of the citation impact environments of scientific journals: An online mapping exercise. \textit{Journal of the American Society for Information Science and Technology} 58(1):25–38.\\
\\
50. 	Leydesdorff L (2008) Caveats for the use of citation indicators in research and journal evaluations. \textit{Journal of the American Society for Information Science and Technology }59(2):278–287.\\
\\
51. 	Schubert A, Glänzel W, Thijs B (2006) The weight of author self-citations. A fractional approach to self-citation counting. \textit{Scientometrics} 67(3):503–514.\\
\\
52. 	Perpiñán Lamigueiro O (2015) \textit{Displaying time series, spatial, and space-time data with R} (CRC Press, Taylor \& Francis Group, Boca Raton, FL).\\
\\
53. 	Garfield E (1972) Citation Analysis as a Tool in Journal Evaluation. Journals can be ranked by frequency and impact of citations for science policy studies. \ 178(4060):471–479.\\
\\
54. 	Nederhof AJ (2006) Bibliometric monitoring of research performance in the Social Sciences and the Humanities: A Review. \textit{Scientometrics} 66(1):81–100.

\end{document}